\documentclass[pra,aps,twocolumn,showpacs]{revtex4}

\usepackage{amsmath}
\usepackage{bm}
\usepackage{graphicx}

\begin{document}

\title{Diagnostics for the ground state phase of a spin-2 Bose-Einstein
condensate}

\author{Hiroki Saito}
\author{Masahito Ueda}
\affiliation{Department of Physics, Tokyo Institute of Technology,
Tokyo 152-8551, Japan \\
and CREST, Japan Science and Technology Corporation (JST), Saitama
332-0012, Japan
}

\date{\today}

\begin{abstract}
We propose a method to determine the singlet-pair energy of a spin-2
Bose-Einstein condensate (BEC).
By preparing the initial populations in the magnetic sublevels $0$ and
$\pm2$ with appropriate relative phases, we can obtain the coefficient
of the spin singlet-pair term from the spin exchange dynamics.
This method is suitable for hyperfine states with short lifetimes, since
only the initial change in the population of each magnetic sublevel is
needed.
This method therefore enables the determination of the ground state phase
of a spin-2 ${}^{87}{\rm Rb}$ BEC at zero magnetic field, which is
considered to lie in the immediate vicinity of the boundary between the
antiferromagnetic and cyclic phases.
We also show that the initial state in which relative phases are
controlled can be prepared by Raman processes.
\end{abstract}


\maketitle

\section{Introduction}

A Bose-Einstein condensate (BEC) in an optical trap~\cite{Kurn} exhibits a
rich variety of spin-related phenomena, such as various magnetic
phases~\cite{Stenger} and spin domain formation~\cite{Miesner}.
The nuclear spin of ${}^{87}{\rm Rb}$ and ${}^{23}{\rm Na}$ is $3 / 2$,
and combined with the spin $1 / 2$ of the outermost electron, the possible
hyperfine spins of these atomic species are $f = 1$ and $f = 2$.
The $f = 1$ BEC was first realized at MIT with ${}^{23}{\rm
Na}$~\cite{Kurn}, for which the antiferromagnetic behavior was
observed~\cite{Stenger}.
On the other hand, ${}^{87}{\rm Rb}$ in the $f = 1$ hyperfine state was
found to be ferromagnetic~\cite{Schmal,Chang}.
The $f = 2$ ${}^{23}{\rm Na}$ condensate was also realized by the MIT
group~\cite{Gorlitz}, whose ground state phase at zero magnetic field is
predicted to be antiferromagnetic~\cite{Ciobanu}.
However, the ground state phase and the spin dynamics of the $f = 2$
${}^{23}{\rm Na}$ BEC have not been studied because of its very short
lifetime (a few milliseconds), due to the fact that the energy of the $f =
2$ state is higher than that of the $f = 1$ state.
The $f = 2$ ${}^{87}{\rm Rb}$ BEC also lies energetically higher than its
$f = 1$ counterpart, but its lifetime is much larger ($\sim 100$ ms)
due to a fortuitous coincidence of the singlet and triplet scattering
lengths~\cite{Julienne}.
The coherent spin dynamics of the $f = 2$ BEC have been
observed~\cite{Schmal,Kuwamoto}.

The ground state of the $f = 2$ BEC of ${}^{87}{\rm Rb}$ is predicted to
be very close to the phase boundary between the antiferromagnetic and
cyclic phases~\cite{Ciobanu,Klausen}.
According to calculations performed by Klausen {\it et
al}.~\cite{Klausen}, the ground state phase of this BEC at zero magnetic
field is barely in the antiferromagnetic phase.
Recent experiments performed by the Hamburg group~\cite{Schmal} and by
the Gakushuin group~\cite{Kuwamoto} appear to support this prediction.
However, due to the experimental uncertainties, the possibility that the
ground state phase is cyclic has not been excluded.
In the Hamburg experiment~\cite{Schmal}, the $m = \pm 2$ mixture was
shown to be stable, which is the ground state of the antiferromagnetic
phase.
However, a magnetic field of 340 mG was applied to the system in this
experiment.
Since the magnetic field lowers the energy of the $m = \pm 2$ states due
to the quadratic Zeeman effect, the observed stability of the $m = \pm 2$
mixture may be due to the presence of the magnetic field.
That there is no spatial separation between the $m = \pm 2$ states
is a necessary condition for a BEC to be antiferromagnetic, but
it is not sufficient.
In the Gakushuin experiment~\cite{Kuwamoto}, the spin dynamics starting
from the $m = 0$ state were investigated for various values of the
magnetic field.
It was found that the population of the $m = \pm 2$ components after 70 ms
evolution increased with a decrease of the magnetic field, which might
imply that the $f = 2$ BEC is antiferromagnetic at zero magnetic field.
However, the initial $m = 0$ state is a highly excited state and the
resultant state after 70 ms is far above the ground state.
Thus, the experimental observations do not exclude the possibility that
the ground state phase at zero magnetic field is cyclic.
In fact, the Hamburg group observed that the spin configuration of the
cyclic phase has a long lifetime~\cite{Schmal,Sengstock}.

There are two experimental difficulties that hinder the determination of
the ground state phase.
One is the short lifetime of $\lesssim$ 100 ms of the upper hyperfine
manifold of ${}^{87}{\rm Rb}$, which makes the equilibrium spin state hard
to reach.
In order to realize the ground state phase, we must equilibrate the system
at least for a few seconds, as in the $f = 1$ case~\cite{Chang,Stenger}.
The other difficulty is the small energy scale ($\sim$ 0.1
nK~\cite{Schmal}) of the spin-singlet-pair term whose sign determines the
ground state phase.
In order to determine the sign of the spin-singlet pair energy, the
magnetic field must be reduced so that the quadratic Zeeman energy is less
than the spin-singlet pair energy.
However, in such a low magnetic field the system may be affected by stray
ac magnetic fields~\cite{Chang}.

To circumvent these difficulties, we propose a method to determine the
value of the spin-singlet pair energy by spin exchange dynamics.
We show that by controlling the initial population and phase in each
magnetic sublevel, we can measure the value of the singlet-pair energy
from the initial spin dynamics.
This method therefore has the advantage that equilibrating the system,
which takes a long time, is not necessary.
Furthermore, the quadratic Zeeman energy is allowed to exceed the
singlet-pair energy and extreme suppression of stray magnetic fields is
not required.

This paper is organized as follows.
Section~\ref{s:bogo} formulates the mean field and Bogoliubov theories of
a spin-2 BEC in the presence of the quadratic Zeeman effect.
Section~\ref{s:homo} studies the spin dynamics in a homogeneous system
and proposes a method to determine the spin-singlet pair energy from the
spin exchange dynamics.
Section~\ref{s:trap} numerically verifies the proposed method for a
trapped system with two-body loss.
Section~\ref{s:exp} discusses the initial spin preparation and
Sec.~\ref{s:conc} concludes the paper.
Detailed derivations of the analytic solutions are given in the Appendix.

\section{Mean field and Bogoliubov analysis for a homogeneous system}
\label{s:bogo}

\subsection{Formulation of the problem}

From rotational symmetry, the low-energy interaction between two atoms
with hyperfine spin $f$ can be classified in terms of the total spin
${\cal F} = 0, 2, \cdots, 2f$, and each scattering channel can be
described by the corresponding $s$-wave scattering length $a_{\cal
F}$~\cite{Ho}.
In the case of $f = 2$ there are three scattering channels ${\cal F} = 0$,
$2$, $4$, and the interaction is described by $(4 \pi \hbar^2 / M)
(a_0 {\cal P}_0 + a_2 {\cal P}_2 + a_4 {\cal P}_4) \delta(\bm{r}_1 -
\bm{r}_2)$, where $M$ is the mass of the atom and ${\cal P}_{\cal F}$
projects the state of two colliding atoms into the state of total spin
${\cal F}$.
The interaction Hamiltonian can be rewritten~\cite{Koashi,Ueda} as $(c_0 +
c_1 \bm{S}_1 \cdot \bm{S}_2 + c_2 {\cal P}_0) \delta(\bm{r}_1 -
\bm{r}_2)$, where $\bm{S}_1$ and $\bm{S}_2$ are the spin operators for
atoms 1 and 2 and the interaction coefficients are given by
\begin{subequations}
\begin{eqnarray}
c_0 & = & \frac{4\pi \hbar^2}{M} \frac{4 a_2 + 3 a_4}{7}, \\
c_1 & = & \frac{4\pi \hbar^2}{M} \frac{a_4 - a_2}{7}, \\
c_2 & = & \frac{4\pi \hbar^2}{M} \frac{7 a_0 - 10 a_2 + 3 a_4}{7}. 
\end{eqnarray}
\end{subequations}

For the linear and quadratic Zeeman effects, the energy of the atom
depends on the magnetic field $B$ as $p m + q m^2$, where $p = \mu_{\rm B}
B / 2$ and $q = -(\mu_{\rm B} B)^2 / (4 \Delta_{\rm hf})$ with $\mu_{\rm
B}$ the Bohr magneton and $\Delta_{\rm hf}$ the hyperfine splitting 
between the states $f = 1$ and $f = 2$~\cite{Pethick}.
We assume $q < 0$ throughout this paper, which is the case for spin-2
${}^{87}{\rm Rb}$.

The mean-field energy of the entire system is given by
\begin{eqnarray} \label{H}
E & = & \int d\bm{r} \Biggl[ \sum_{m = -2}^2 \psi_m^*
\left( -\frac{\hbar^2}{2M} \nabla^2 + V + p m + q m^2 \right) \psi_m
\nonumber \\
& & + \frac{c_0}{2} n_{\rm tot}^2 + \frac{c_1}{2} \bm{F}^2 + \frac{c_2}{2}
|A_{00}|^2 \Biggr],
\end{eqnarray}
where $m = -2, \cdots, 2$ denotes the magnetic sublevels, $V$ is an
external potential,
\begin{equation}
n_{\rm tot} = \sum_{m = -2}^{2} |\psi_m|^2
\end{equation}
is the total particle-number density,
\begin{equation}
\bm{F} = \sum_{m m'} \psi_m^* \bm{S}_{m m'} \psi_{m'}
\end{equation}
is the spin vector with $5 \times 5$ spin-2 matrix $\bm{S}$, and
\begin{equation}
A_{00} = \frac{1}{\sqrt{5}} \left( 2 \psi_2 \psi_{-2} - 2 \psi_1 \psi_{-1}
+ \psi_0^2 \right)
\end{equation}
is the singlet-pair amplitude.
Minimizing the action,
\begin{equation}
S = \int dt (i \hbar \sum_m \psi_m^* \partial_t \psi_m - E),
\end{equation}
gives the multicomponent Gross-Pitaevskii (GP) equations for the
spin-2 BEC as
\begin{subequations}
\label{GP}
\begin{eqnarray} 
i \hbar \frac{\partial \psi_{\pm 2}}{\partial t} & = &
\left(-\frac{\hbar^2}{2M} \nabla^2 + V \pm 2 p + 4 q \right) \psi_{\pm 2}
+ c_0 n_{\rm tot} \psi_{\pm 2} \nonumber \\
& & + c_1 \left( F_{\mp} \psi_{\pm 1} \pm 2 F_z
\psi_{\pm 2} \right) + \frac{c_2}{\sqrt{5}} A_{00} \psi_{\mp 2}^*,
\label{GP1} \\
i \hbar \frac{\partial \psi_{\pm 1}}{\partial t} & = &
\left(-\frac{\hbar^2}{2M} \nabla^2 + V \pm p + q \right) \psi_{\pm 1} +
c_0 n_{\rm tot} \psi_{\pm 1} \nonumber \\
& & + c_1 \left( \frac{\sqrt{6}}{2} F_{\mp}
\psi_0 + F_{\pm} \psi_{\pm 2} \pm F_z \psi_{\pm 1} \right) \nonumber \\
&& - \frac{c_2}{\sqrt{5}} A_{00} \psi_{\mp 1}^*, \label{GP2}
\\
i \hbar \frac{\partial \psi_0}{\partial t} & = & \left(-\frac{\hbar^2}{2M}
\nabla^2 + V \right) \psi_0 + c_0 n_{\rm tot} \psi_0 \nonumber \\
& & + \frac{\sqrt{6}}{2}
c_1 \left( F_- \psi_{-1} + F_+ \psi_1 \right) + \frac{c_2}{\sqrt{5}}
A_{00} \psi_{0}^*, \label{GP3}
\end{eqnarray}
\end{subequations}
where $F_\pm = F_x \pm i F_y$.

We note that the linear Zeeman term only rotates the hyperfine spin about
the $z$ axis, and hence the spin dynamics are essentially independent of
the linear Zeeman term.
In fact, setting $\psi_m \rightarrow e^{i p m t / \hbar} \psi_m$ and
noting that the phase factor also arises in the $F_\pm$ terms as $e^{\mp i
p t / \hbar} F_\pm$, we can completely eliminate the linear Zeeman terms
in Eqs.~(\ref{GP1}) and (\ref{GP2}).
This property is due to the rotational symmetry of our system with
respect to the $z$ axis, which results in the conservation of the
projected angular momentum on the $z$ axis, $\langle F_z \rangle = \int
d\bm{r} \sum_m m |\psi_m|^2$.

\subsection{Ground states in a homogeneous system}

In an ultracold spinor BEC which is isolated from the environment, the
total spin angular momentum in the direction of the magnetic field is
conserved for a long time ($\gtrsim$ 1 sec)~\cite{Chang}.
We therefore minimize the energy of the system in the subspace of a given
$\langle F_z \rangle = \int d\bm{r} \sum_m m |\psi_m|^2$; we shall refer
to the resulting minimized state as the ``ground state''.
For simplicity, we restrict ourselves to the subspace of $\langle F_z
\rangle = 0$ in this paper, i.e., we seek the ground state in the subspace
of $\langle F_z \rangle = 0$.

We consider the uniform case by setting $V = 0$ and $\psi_m = \sqrt{n}
\zeta_m$, where the density $n$ is constant and $\zeta_m$ satisfies
$\sum_m |\zeta_m|^2 = 1$.
The energy per atom is given by
\begin{equation} \label{ene}
\varepsilon \equiv \frac{E}{N} = q \sum_{m = -2}^{2} m^2 |\zeta_m|^2 +
\frac{\tilde c_0}{2} + \frac{\tilde c_1}{2} \bm{f}^2 + \frac{\tilde
c_2}{2} |a_{00}|^2,
\end{equation}
where $\bm{f} \equiv \bm{F} / n = \sum_{mm'} \zeta_m^* \bm{S}_{mm'}
\zeta_{m'}$, $a_{00} = A_{00} / n = (2 \zeta_2 \zeta_{-2} - 2 \zeta_1
\zeta_{-1} + \zeta_0^2) / \sqrt{5}$, and
\begin{equation}
\tilde c_i \equiv c_i n
\end{equation}
for $i = 0$, $1$, $2$.
The linear Zeeman term is absent in Eq.~(\ref{ene}) because $\langle F_z
\rangle = 0$.

In order to find the ground state, we compare the energies of the
stationary states given in Ref.~\cite{Sengstock}.
The energy of the antiferromagnetic state,
\begin{equation}
\bm{\zeta}_{\rm AF} = (e^{i \chi_2}, 0, 0, 0, e^{i \chi_{-2}}),
\end{equation}
where $\chi_m$ is the phase of each component, is given by
\begin{equation}
\varepsilon_{\rm AF} = 4q + \tilde c_0 / 2 + \tilde c_2 / 10.
\end{equation}
This energy is independent of the phase factor $e^{i \chi_m}$.
For the cyclic state,
\begin{equation} \label{zeta}
\bm{\zeta}_{\rm C} = \left( \frac{1}{\sqrt{2}} e^{i \chi_2} \sin\theta, 0,
e^{i \chi_0} \cos\theta, 0, \frac{1}{\sqrt{2}} e^{i \chi_{-2}} \sin\theta
\right),
\end{equation}
the energy of the system takes the form
\begin{equation} \label{EN}
\varepsilon_{\rm C} = \frac{\tilde c_0}{2} + \frac{\tilde c_2}{10} \left|
\cos^2\theta + e^{i \chi} \sin^2\theta \right|^2 + 4 q \sin^2\theta,
\end{equation}
where $\chi \equiv \chi_2 + \chi_{-2} - 2\chi_0$.
When $\tilde c_2 < 10 |q|$, Eq.~(\ref{EN}) is minimized by $\theta = \pi /
2$, which corresponds to the antiferromagnetic state $\bm{\zeta}_{\rm AF}$.
When $\tilde c_2 > 10 |q|$, Eq.~(\ref{EN}) becomes minimal with $\chi =
\pi$ and 
\begin{equation} \label{cyclic}
\cos^2\theta = \frac{1}{2} + \frac{5 q}{\tilde c_2}.
\end{equation}
Substituting this into Eq.~(\ref{EN}), we obtain
\begin{equation}
\varepsilon_{\rm C} = 2q + \frac{\tilde c_0}{2} - \frac{10 q^2}{\tilde
c_2} \;\;\;\;\;\;\;\;\; (\tilde c_2 > 10 |q|).
\end{equation}
The states $\bm{\zeta}_{\rm AF}$ and $\bm{\zeta}_{\rm C}$ satisfy
$F_z(\bm{r}) = \sum_m m |\psi_m(\bm{r})|^2 = 0$.

Even if the local spin $F_z(\bm{r})$ is nonzero, the total spin $\langle
F_z \rangle$ can be zero by the formation of the spatial spin
structure~\cite{Saito}.
For example, a staggered domain structure of the ferromagnetic state
gives $\langle F_z \rangle = 0$.
The kinetic energy at the domain wall can be neglected when the region of
the wall is much smaller than the region of the domain.
Hence, we also consider the case of $F_z(\bm{r}) \neq 0$, bearing in mind
that the total spin is maintained as zero, $\langle F_z \rangle = 0$, by
some structure formation.
The energy of the ferromagnetic state, $\bm{\zeta}_{\rm F} = (e^{i
\chi_2}, 0, 0, 0, 0)$ or $(0, 0, 0, 0, e^{i \chi_{-2}})$, is given by
\begin{equation}
\varepsilon_{\rm F} = 4q + \tilde c_0 / 2 + 2 \tilde c_1.
\end{equation}
We also consider a state found in Ref.~\cite{Sengstock},
\begin{eqnarray}
\bm{\zeta}_{\rm M} & = & (e^{i \chi_2} \cos\theta, 0, 0, e^{i \chi_{-1}}
\sin\theta, 0) \nonumber \\
& \mbox{or} & (0, e^{i \chi_1} \sin\theta, 0, 0, e^{i \chi_{-2}}
\cos\theta).
\end{eqnarray}
The energy of this state is obtained to be
\begin{equation} \label{es}
\varepsilon_{\rm M} = q (1 + 4 \cos^2 \theta) + \frac{\tilde c_0}{2} +
\frac{\tilde c_1}{2} (3 \cos^2 \theta - 1)^2.
\end{equation}
This is minimized by $\cos^2\theta = 1 / 3 - q / (3 \tilde c_1)$ for
$\tilde c_1 > |q| / 2$, and the energy then becomes
\begin{equation}
\varepsilon_{\rm M} = 2 q + \frac{\tilde c_0}{2} - \frac{q^2}{2 \tilde
c_1} \;\;\;\;\;\;\;\;\; (\tilde c_1 > |q| / 2).
\end{equation}

\begin{figure}[tb]
\includegraphics[width=8.4cm]{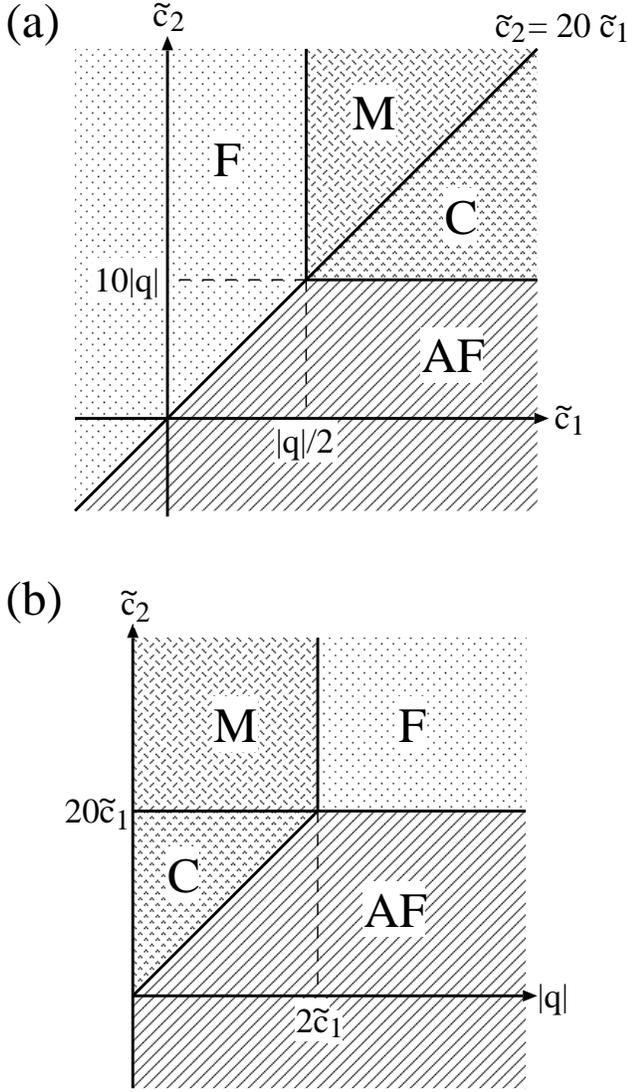}
\caption{
The ground-state phase diagrams for $p = 0$ and $q < 0$ (a) with respect
to $\tilde c_1$ and $\tilde c_2$ and (b) with respect to $|q|$ and $\tilde
c_1$.
The symbols AF, C, F, and M indicate the states $\bm{\zeta}_{\rm AF}$,
$\bm{\zeta}_{\rm C}$, $\bm{\zeta}_{\rm F}$, and $\bm{\zeta}_{\rm M}$,
respectively.
In (b), $\tilde c_1 > 0$ is assumed.
}
\label{f:diagram}
\end{figure}

Comparing the energies $\varepsilon_{\rm AF}$, $\varepsilon_{\rm F}$,
$\varepsilon_{\rm C}$, and $\varepsilon_{\rm M}$, we obtain the phase
diagrams shown in Fig.~\ref{f:diagram}.
We have assumed here that one of the states $\bm{\zeta}_{\rm AF}$,
$\bm{\zeta}_{\rm C}$, $\bm{\zeta}_{\rm F}$, and $\bm{\zeta}_{\rm M}$ is
the ground state, based on the numerical results of
Ref.~\cite{Sengstock}.
We have confirmed that this assumption is correct using the Monte Carlo
method.

In the case of ${}^{87}{\rm Rb}$, $c_1$ is positive and $|c_2|$ is at most
of the same order of magnitude as $c_1$~\cite{Klausen}.
Hence, the ground state of the spin-2 ${}^{87}{\rm Rb}$ BEC is
antiferromagnetic or cyclic depending on whether $\tilde c_2 < 10 |q|$ or
$\tilde c_2 > 10 |q|$.
Therefore, in order to determine the sign of $c_2$ from the ground state,
we must suppress the magnetic field so that the condition $|q| < \tilde
c_2 / 10$ is met.
In the experiment of the Hamburg group~\cite{Schmal}, a magnetic field
of $B = 340$ mG was applied, which corresponds to $|q| / k_{\rm B} \simeq
0.4$ nK.
The mean atomic density was $n \simeq 4 \times 10^{14} {\rm cm}^{-3}$, and
if $|c_2| M / (4\pi \hbar^2) \simeq a_{\rm B}$, we have $|\tilde c_2| /
k_{\rm B} \simeq 1.5$ nK.
Thus, $\tilde c_2 \lesssim 10 |q|$, which is consistent with the observed
stability of the antiferromagnetic state $\bm{\zeta}_{\rm AF}$.
In the Gakushuin experiment~\cite{Kuwamoto}, $B > 100$ mG and $n \simeq
2.1 \times 10^{14} {\rm cm}^{-3}$, corresponding to $|q| \gtrsim 0.03$ nK
and $|\tilde c_2| / k_{\rm B} \simeq 0.8$ nK.
Hence, the reported parameters of Ref.~\cite{Kuwamoto} belong to the
cyclic phase.

\subsection{Stability of the stationary states}

The antiferromagnetic (or stretched) state,
\begin{equation} \label{polar}
\bm{\Psi}_{\rm AF} = \sqrt{\frac{n}{2}} (1, 0, 0, 0, 1),
\end{equation}
was found in experiments to be stable in the presence of a magnetic
field~\cite{Schmal}.
To examine this result, we will investigate the stability of this state
using Bogoliubov analysis.

Substituting Eq.~(\ref{polar}) into Eqs.~(\ref{GP}), we find that it
evolves as $e^{-i \mu t / \hbar} (e^{-2 i p t / \hbar}, 0, 0, 0, e^{2 i p
t / \hbar})$ with $\mu = 4 q + \tilde c_0 + \tilde c_2 / 5$, and hence
Eq.~(\ref{polar}) is stationary in the rotating frame $e^{i m p} \psi_m$.
We can therefore perform Bogoliubov analysis with respect to the state
(\ref{polar}) in the rotating frame, which is described by the GP
equations (\ref{GP}) without the linear Zeeman terms.
We set the wave function as $\bm{\psi} = e^{-i \mu t / \hbar}
(\bm{\Psi}_{\rm AF} + \bm{\phi})$ and substitute it into the GP equations
in the rotating frame.
Taking the linear terms with respect to $\bm{\phi}$, we obtain
\begin{subequations} \label{Bogo}
\begin{eqnarray}
i \hbar \frac{\partial \phi_{\pm 2}}{\partial t} & = & -\frac{\hbar^2}{2M}
\nabla^2 \phi_{\pm 2} + \left( \frac{\tilde c_0}{2} + 2 \tilde c_1 \right)
(\phi_{\pm 2} + \phi_{\pm 2}^*) \nonumber \\
& & + \left( \frac{\tilde c_0}{2} - 2 \tilde
c_1 + \frac{\tilde c_2}{5} \right) (\hat\phi_{\mp 2} + \phi_{\mp 2}^*),
\\
i \hbar \frac{\partial \phi_{\pm 1}}{\partial t} & = & \left(
-\frac{\hbar^2}{2M} \nabla^2 - 3q \right) \phi_{\pm 1} \nonumber \\
& & + \left( \tilde c_1
- \frac{\tilde c_2}{5} \right) (\phi_{\pm 1} + \hat\phi_{\mp 1}^*), \\
i \hbar \frac{\partial \phi_0}{\partial t} & = & \left(
-\frac{\hbar^2}{2M} \nabla^2 - 4q - \frac{\tilde c_2}{5} \right) \phi_0 +
\frac{\tilde c_2}{5} \phi_0^*,
\end{eqnarray}
\end{subequations}
These equations can be reduced to the eigenvalue problem by expansion
of $\bm{\phi}$ as
\begin{equation}
\bm{\phi}(\bm{r}, t) = \sum_{\bm{k}} \left( \bm{u}_{\bm{k}} e^{i
(\bm{k} \cdot \bm{r} - \omega_k t)} + \bm{v}_{\bm{k}}^* e^{-i
(\bm{k} \cdot \bm{r} - \omega_k t)} \right),
\end{equation}
and the eigenenergies are obtained as
\begin{subequations} \label{eigen}
\begin{eqnarray}
& & \left[ \varepsilon_k \left( \varepsilon_k + 2 \tilde c_0 + 2 \tilde
c_2 / 5 \right) \right]^{1/2}, \label{mode1} \\
& & \left[ \varepsilon_k \left( \varepsilon_k + 8 \tilde c_1 - 2 \tilde
c_2 / 5 \right) \right]^{1/2}, \label{mode2} \\
& & \left[ (\varepsilon_k - 3q) \left( \varepsilon_k - 3q + 2 \tilde c_1 -
2 \tilde c_2 / 5 \right) \right]^{1/2}, \label{mode3} \\
& & \left[ (\varepsilon_k - 4q) \left( \varepsilon_k - 4q - 2 \tilde c_2 /
5 \right) \right]^{1/2}, \label{mode4}
\end{eqnarray}
\end{subequations}
where $\varepsilon_k \equiv (\hbar k)^2 / (2M)$.

If the eigenenergy is complex, the corresponding mode is dynamically
unstable against exponential growth.
The eigenvectors $\bm{u}_{\bm k}$ and $\bm{v}_{\bm k}$ of the first
mode~(\ref{mode1}) are proportional to $(1, 0, 0, 0, 1)$, which is the
same form as $\bm{\Psi}_{\rm AF}$, indicating that a collapse occurs if
$2 \tilde c_0 + 2 \tilde c_2 / 5 < 0$.
The eigenvectors of the second mode~(\ref{mode2}) are proportional to $(1,
0, 0, 0, -1)$.
Since this eigenmode transfers the $m = \pm 2$ component to the $m = \mp
2$ component, excitation of the mode~(\ref{mode2}) gives rise to exchange
of atoms between the $m = \pm 2$ components.
This implies that phase separation between the two components occurs
for $8 \tilde c_1 - 2 \tilde c_2 / 5 < 0$.
Therefore, the fact that no phase separation is observed in
experiments~\cite{Schmal} only indicates that $\tilde c_2 < 20 \tilde
c_1$; it is not sufficient to conclude that $\tilde c_2 < 0$.
The third mode~(\ref{mode3}) is two-fold degenerate and the eigenvectors
are proportional to $(0, 1, 0, 0, 0)$ and $(0, 0, 0, 1, 0)$.
Therefore, when $2 \tilde c_1 - 2 \tilde c_2 / 5 < 3q$, the state
(\ref{polar}) is dynamically unstable against the growth of the $m = \pm
1$ components.
Similarly the last mode~(\ref{mode4}) is proportional to $(0, 0, 1, 0, 0)$
and describes the growth of the $m = 0$ component.
Dynamical instability arises in this mode for $\tilde c_2 > 10 |q|$,
consistent with the fact that the ground state has the $m = 0$
component for $\tilde c_2 > 10 |q|$, as shown by Eq.~(\ref{cyclic}).

For later use, we also perform Bogoliubov analysis for the stationary
state $\bm{\Psi} = (0, 0, \sqrt{n}, 0, 0)$.
The eigenenergies are obtained to be
\begin{subequations} \label{eigen2}
\begin{eqnarray}
& & \left[ \varepsilon_k \left( \varepsilon_k + 2 \tilde c_0 + 2 \tilde
c_2 / 5 \right) \right]^{1/2}, \label{mode2_1} \\
& & \left[ (\varepsilon_k + q) \left( \varepsilon_k + q + 6 \tilde c_1 -
2 \tilde c_2 / 5 \right) \right]^{1/2}, \label{mode2_2} \\
& & \left[ (\varepsilon_k + 4q) \left( \varepsilon_k + 4q - 2 \tilde c_2 /
5 \right) \right]^{1/2}. \label{mode2_3}
\end{eqnarray}
\end{subequations}
The eigenvector of the first mode (\ref{mode2_1}) is proportional to $(0,
0, 1, 0, 0) \propto \bm{\Psi}$, implying that the state $\bm{\Psi}$
collapses if $\tilde 2 c_0 + 2 \tilde c_2 / 5 < 0$.
We note that this condition for the collapse of the BEC is the same as
that for $\bm{\Psi}_{\rm AF}$.
The second and third modes, Eqs.~(\ref{mode2_2}) and (\ref{mode2_3}), are
both two-fold degenerate.
The eigenvectors of the mode (\ref{mode2_2}) are proportional to $(0, 1,
0, 0, 0)$ and $(0, 0, 0, 1, 0)$, and therefore the dynamical instability
in this mode increases the $m = \pm 1$ components.
Similarly, the mode (\ref{mode2_3}) has eigenvectors proportional to
$(1, 0, 0, 0, 0)$ and $(0, 0, 0, 0, 1)$.
In contrast to the excitations for $\bm{\Psi}_{\rm AF}$, there are
always dynamically unstable modes in Eqs.~(\ref{mode2_2}) and
(\ref{mode2_3}) because of $q < 0$, and therefore the $m = 0$ state is
always dynamically unstable for $B > 0$ in a homogeneous system.

\section{Spin dynamics in a homogeneous system}
\label{s:homo}

\subsection{Analytic solutions}

In order to determine the value of $c_2$ from the spin dynamics, we
consider the spin dependent interaction that is sensitive only to the
spin-singlet term.
The elementary processes in the spin-singlet channel are $0 + 0
\leftrightarrow 0 + 0$, $0 + 0 \leftrightarrow 1 + (-1)$, $0 + 0
\leftrightarrow 2 + (-2)$, and $1 + (-1) \leftrightarrow 2 + (-2)$.
On the other hand, the $c_1$ term consists of elementary processes
such as $m_1 + m_2 \leftrightarrow (m_1 + 1) + (m_2 - 1)$.
The spin independent interaction does not flip the spin.
Thus, the elementary process appearing only in the $c_2$ term is $0 + 0
\leftrightarrow 2 + (-2)$.
We therefore focus on this process and consider the initial state of the
form
\begin{equation} \label{initz}
\bm{\zeta} = \left( \frac{1}{\sqrt{2}} e^{i \chi_2} \sin\theta, 0,
e^{i \chi_0} \cos\theta, 0, \frac{1}{\sqrt{2}} e^{i \chi_{-2}} \sin\theta
\right).
\end{equation}
Since the overall phase and spin rotation about the $z$ axis do not
affect the physics, the relevant phase appears only in the combination of
$\chi_2 + \chi_{-2} - 2\chi_0$.
We therefore assume that $\chi_0 = \chi_2 = 0$ and $0 \leq \theta \leq \pi
/ 2$ without loss of generality.

If $\zeta_{\pm 1}$ are exactly zero in the initial state, we find from
Eq.~(\ref{GP2}) that $\zeta_{\pm 1}(t)$ are always zero within the mean
field approximation.
We therefore assume $\zeta_{\pm 1}(t) = 0$ in the following analysis.
In experiments~\cite{Schmal,Sengstock}, this assumption holds at least
for an initial period of $\sim 100$ ms.
Although the $m = \pm 1$ components exponentially grow in the presence of
the dynamical instability, the assumption $\zeta_{\pm 1}(t) = 0$ is still
valid in the early stage of time evolution.

The GP equations then reduce to
\begin{subequations} \label{GP02}
\begin{eqnarray}
i \hbar \dot\zeta_0 & = & \frac{\tilde c_2}{5} \zeta_0^* (2 \zeta_2
\zeta_{-2} + \zeta_0^2), \label{GP02_0} \\
i \hbar \dot\zeta_{\pm 2} & = & \frac{\tilde c_2}{5} \zeta_{\mp 2}^* (2
\zeta_2 \zeta_{-2} + \zeta_0^2) + 4 q \zeta_{\pm 2}. \label{GP02_1}
\end{eqnarray}
\end{subequations}
Differentiating Eq.~(\ref{GP02_0}) with respect to time and using
Eq.~(\ref{GP02_1}), we can eliminate $\zeta_{\pm 2}$ to obtain
\begin{equation} \label{eqzeta0}
\hbar^2 \ddot\zeta_0 = \frac{2 \tilde c_2}{5} \left( \varepsilon - 4 q +
8 q |\zeta_0|^2 \right) \zeta_0 - i \left( \frac{2 \tilde c_2}{5} + 8 q
\right) \hbar \dot\zeta_0,
\end{equation}
where the energy per atom,
\begin{equation} \label{E1}
\varepsilon = \frac{\tilde c_2}{10} \left|2 \zeta_2 \zeta_{-2} + \zeta_0^2
\right|^2 + 4 q \left( |\zeta_2|^2 + |\zeta_{-2}|^2 \right),
\end{equation}
is a constant of motion.
It is interesting to note that the form of Eq.~(\ref{eqzeta0}) coincides
with that describing a 1D BEC on a rotating ring if the time derivative is
replaced with the spatial derivative~\cite{Kanamoto}.
The solution is obtained (see Appendix for derivation) as
\begin{equation} \label{SOL1}
\zeta_0(t) = e^{i \varphi_0(t)} \sqrt{A_0 {\rm sn}^2(\alpha t + \beta_0 |
\nu) + B_0},
\end{equation}
where $\varphi_0(t)$, $A_0$, $B_0$, $\alpha$, $\beta_0$, and $\nu$ are
given by Eqs.~(\ref{phi0}), (\ref{a0org}), (\ref{eqb0}), (\ref{alphorg}),
(\ref{beta0org}), and (\ref{nuorg}), respectively.
The function sn is a Jacobian elliptic function~\cite{Handbook}.

If we assume $\chi_{-2} = \pi$ in the initial state, i.e.,
\begin{equation} \label{init}
\bm{\zeta}(0) = \left( \frac{1}{\sqrt{2}} \sin\theta_0, 0, \cos\theta_0,
0, -\frac{1}{\sqrt{2}} \sin\theta_0 \right),
\end{equation}
the constants in the solution (\ref{SOL1}) become $\beta_0 = 0$, $B_0 =
\zeta_0^2(0) = \cos^2\theta_0$, and
\begin{eqnarray}
A_0 & = & \frac{\tilde c_2}{10 q} \zeta_0^2(0) \left[ 1 -
\zeta_0^2(0) \right], \label{A0} \\
\alpha & = & \sqrt{\frac{8q}{5 \hbar^2} \left\{ \tilde c_2 \left[ 1 - 2
\zeta_0^2(0) \right] + 10 q \right\} }, \label{alpha} \\
\nu & = & \frac{\tilde c_2^2 \zeta_0^2(0) \left[ 1 - \zeta_0^2(0)
\right]}{10 q \left\{ \tilde c_2 \left[ 1 - 2 \zeta_0^2(0) \right] + 10 q
\right\}}. \label{m}
\end{eqnarray}
The sign of $A_0$ is opposite to that of $c_2$ because $q < 0$.
The constants $\alpha$ and $\nu$ are real and positive if $\tilde c_2
\left[ 1 - 2 \zeta_0^2(0) \right] + 10 q < 0$.

\subsection{Measurement of $c_2$}
\label{ss:main}

We suppose that the initial state (\ref{init}) is prepared and the
magnetic field satisfies $|q| > |\tilde c_2| / 10$.
Under this condition, we find that $\alpha$ and $\nu$ in
Eqs.~(\ref{alpha}) and (\ref{m}) are real and positive.
Since the atomic density is typically $n \sim 10^{14} {\rm cm}^{-3}$ in
experiments~\cite{Schmal,Kuwamoto} and $|(7 a_0 - 10 a_2 
+ 3 a_4) / 7|$ is of order of the Bohr radius at most~\cite{Klausen}, the
condition $|q| > |\tilde c_2| / 10$ requires $B \gtrsim 100$ mG.
At this magnetic field, the ground state in the subspace of $\langle F_z
\rangle = 0$ is the stretched state (\ref{polar}) irrespective of the sign
of $c_2$.

The Jacobian elliptic function ${\rm sn}(\alpha t | \nu)$ initially
increases for positive $\alpha$ and $\nu$, and hence the $m = 0$ component
$|\zeta_0(t)|^2$ in Eq.~(\ref{SOL1}) initially decreases for $c_2 > 0$ and
increases for $c_2 < 0$.
Therefore, we can determine the sign of $c_2$ from the sign of the initial
change in the spin population, which is the main result of this paper.
This method is suitable for atomic species with short lifetimes, since
only the initial stage of time evolution is needed for the determination
of $c_2$.

\begin{figure}[tb]
\includegraphics[width=8.4cm]{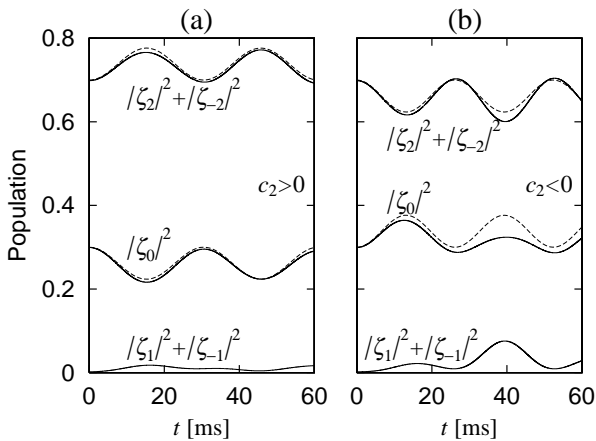}
\caption{
Time evolution of the relative population of each spin component for a
density of atoms of $2.1 \times 10^{14} {\rm
cm}^{-3}$~\protect\cite{Kuwamoto} and a magnetic field of 250 mG.
The scattering lengths are assumed to be $(c_1, c_2) M / (4 \pi \hbar^2) =
(a_{\rm B}, a_{\rm B})$ in (a) and $(a_{\rm B}, -a_{\rm B})$ in (b), where
$a_{\rm B}$ is the Bohr radius.
The initial state is chosen to be $\zeta_0 = \sqrt{0.3}$, $\zeta_2 =
-\zeta_{-2} = \sqrt{0.35}$, and $\zeta_{\pm 1} = \sqrt{0.001}$.
The dashed curves are drawn according to Eq.~(\protect\ref{sin}).
}
\label{f:unifev}
\end{figure}

Figure~\ref{f:unifev} shows time evolution of the spin populations which
are obtained by numerically solving Eq.~(\ref{GP02}).
The results confirm the relation between the sign of $c_2$ and the sign of
the initial change in the spin populations.
In Fig.~\ref{f:unifev}, the $m = \pm 1$ components are assumed to have
small initial values ($\zeta_{\pm 1} = \sqrt{0.001}$) to assess imperfect
population transfer in realistic situations.
We find that the dynamical instability in the $m = \pm 1$ components
have little effect on the initial dynamics of the $m = 0$, $\pm 2$
components.

In Fig.~\ref{f:unifev}, $\nu \sim 0.01$.
For $0 \leq \nu \ll 1$, the Jacobian elliptic function can be approximated
by the trigonometric function as~\cite{Handbook}
\begin{equation}
{\rm sn}(\alpha t | \nu) = \sin \alpha t + O(\nu),
\end{equation}
and the spin evolution is then approximated by
\begin{subequations} \label{sin}
\begin{eqnarray}
|\zeta_0(t)|^2 & \simeq & \cos^2 \theta_0 + A_0 \sin^2 \alpha t, \\
|\zeta_{\pm 2}(t)|^2 & \simeq & \frac{1}{2} \left( \sin^2 \theta_0 - A_0
\sin^2 \alpha t \right).
\end{eqnarray}
\end{subequations}
The amplitude of this oscillation is $A_0 \propto -c_2$, showing
explicitly that whether the spin population initially increases or
decreases depends on the sign of $c_2$.
We can also determine the magnitude of $c_2$ by measuring the amplitude of
the oscillations.
In Fig.~\ref{f:unifev}, Eq.~(\ref{sin}) is plotted as dashed curves,
which are in good agreement with the numerical solutions (solid curves) in
the early stages of the time evolution.

The differences between the numerical and analytic curves in
Fig.~\ref{f:unifev} arise mainly from the growth of the $m = \pm 1$
components in the numerical result due to the dynamical instability.
We find that the growth of the $m = \pm 1$ components in
Fig.~\ref{f:unifev} (b) is larger than that in Fig.~\ref{f:unifev} (a).
This can qualitatively be understood from the Bogoliubov energy
(\ref{mode2_2}), which describes the growth of the $m = \pm 1$ components
from the state $\bm{\Psi} = (0, 0, \sqrt{n}, 0, 0)$.
The imaginary part of Eq.~(\ref{mode2_2}) at $\varepsilon_k = 0$, $[q (q +
6 \tilde c_1 - 2 \tilde c_2 / 5)]^{1/2}$, is larger for $c_2 < 0$ than for
$c_2 > 0$ since $q < 0$ and $q + 6 \tilde c_1 > 0$ in the present case.

\begin{figure}[tb]
\includegraphics[width=8.4cm]{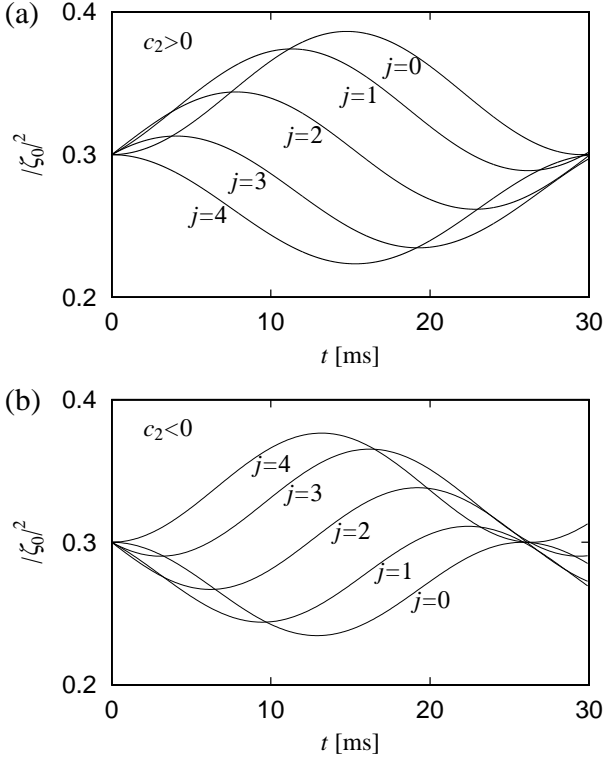}
\caption{
Time evolution of $|\zeta_0|^2$ with the same parameters as in
Figs.~\protect\ref{f:unifev} (a) and (b).
The initial state is given by $\zeta_0 = \sqrt{0.3}$, $\zeta_{\pm 1} = 0$,
$\zeta_2 = \sqrt{0.35}$, and $\zeta_{-2} = \exp(i \pi j / 4) \sqrt{0.35}$,
where $j$ is an integer ranging from 0 to 4.
}
\label{f:phase}
\end{figure}

We note that the relative phase $\chi \equiv \chi_2 + \chi_{-2} - 2\chi_0
= \pi$ in the initial state (\ref{initz}) is crucial for determining the
value of $c_2$ by the above method.
For example, when the initial state is given by Eq.~(\ref{initz}) with
$\chi = 0$, or without loss of generality,
\begin{equation} \label{sameph}
\bm{\zeta} = \left( \frac{1}{\sqrt{2}} \sin\theta_0, 0, \cos\theta_0,
0, \frac{1}{\sqrt{2}} \sin\theta_0 \right),
\end{equation}
the time evolution becomes Eq.~(\ref{SOL1}), where $A_0$, $\nu$, and
$\alpha$ are given by Eqs.~(\ref{samea0}), (\ref{samenu}), and
(\ref{samealph}), respectively (see Appendix for details).
These constants satisfy $c_2 A_0 > 0$ and $\nu < 0$, and $\alpha$ is real.
Using the relation~\cite{Handbook}
\begin{equation}
{\rm sn}(\alpha t \bigm| -|\nu|) = \frac{1}{\sqrt{1 + |\nu|}} {\rm sd}
\left( \sqrt{1 + |\nu|} \alpha t \biggm| \frac{|\nu|}{1 + |\nu|} \right)
\end{equation}
and noting that the Jacobian elliptic function ${\rm sd} = {\rm sn} / {\rm
dn}$ initially increases, we find that $|\zeta_0(t)|^2$ initially
decreases for $c_2 < 0$ and increases for $c_2 > 0$.
This behavior is opposite to that in Fig.~\ref{f:unifev} and therefore
control of the relative phase $\chi$ in the initial state is crucial
for determining the sign of $c_2$.
Figure~\ref{f:phase} shows the initial evolution of $|\zeta_0|^2$ with
$\chi$ varying from $0$ to $\pi$, which indicates that the phase of the
oscillation depends on $\chi$.

\subsection{Dynamics in the cyclic phase}

If $|q| < \tilde c_2 / 10$, the stretched state (\ref{polar}) is
dynamically unstable against growth of the $m = 0$ component, as shown by
Eq.~(\ref{mode4}).
This is because the ground state is in the cyclic phase given by
Eq.~(\ref{zeta}) if $|q| < \tilde c_2 / 10$ and $\tilde c_2 < 20 \tilde
c_1$.
We can therefore conclude that $\tilde c_2 - 10 |q| > 0$ and hence $c_2 >
0$, if we observe the growth of the $m = 0$ component from the initial
stretched state.

\begin{figure}[tb]
\includegraphics[width=8.4cm]{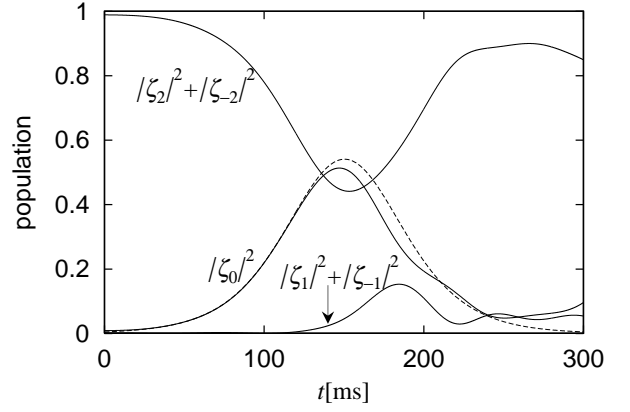}
\caption{
Time evolution of the spin populations with a magnetic field of 100 mG.
The other parameters are the same as in Fig.~\protect\ref{f:unifev} (a).
The initial state is given by $\zeta_0 = \sqrt{0.009}$, $\zeta_{\pm 1} =
\sqrt{0.001}$, $\zeta_2 = -\zeta_{-2} = \sqrt{0.99}$.
The dashed curve is drawn according to Eq.~(\protect\ref{tanh}).
}
\label{f:cyclic}
\end{figure}

Suppose that $\cos^2 \theta_0 \ll 1$ in the initial state given by
Eq.~(\ref{init}) and $|q| < \tilde c_2  / 10$, then $\nu$ is negative and
$\alpha$ is pure imaginary from Eqs.~(\ref{alpha}) and (\ref{m}).
In this case, the Jacobian elliptic function can be rewritten as
\begin{equation} \label{dncd}
{\rm sn}(i |\alpha| t \bigm| -|\nu|) = \frac{i}{\sqrt{1 + |\nu|}} {\rm sd}
\left( \sqrt{1 + |\nu|} |\alpha| t \biggm| \frac{1}{1 + |\nu|} \right).
\end{equation}
Since $|\nu| \ll 1$ and hence $1 / (1 + |\nu|) \simeq 1$, we can make the
approximation~\cite{Handbook}
\begin{equation} \label{sdeq}
{\rm sd}(u | (1 + |\nu|)^{-1}) \simeq \frac{\sinh u + \frac{|\nu|}{4}
(\sinh u \cosh u - u) {\rm sech} u}{1 + \frac{|\nu|}{4} (\sinh u \cosh u -
u) \tanh u},
\end{equation}
where $u \equiv |\alpha| t$.
When $u \gg 1$, Eq.~(\ref{sdeq}) reduces to
\begin{equation}
{\rm sd}(u | (1 + |\nu|)^{-1}) \simeq \frac{1}{\sqrt{|\nu|}} {\rm sech}
\left( |\alpha| t - \frac{1}{2} \ln \frac{16}{|\nu|} \right).
\end{equation}
Equation~(\ref{SOL1}) is thus approximated to give
\begin{equation} \label{tanh}
|\zeta_0(t)|^2 \simeq -\frac{A_0}{|\nu|} {\rm sech}^2\left( |\alpha| t -
\frac{1}{2} \ln \frac{16}{|\nu|} \right) + \cos^2\theta_0.
\end{equation}
The dashed curve in Fig.~\ref{f:cyclic} shows Eq.~(\ref{tanh}).
A comparison of this with the corresponding numerical result (solid
curve) shows that Eq.~(\ref{tanh}) is in good agreement with the numerical
result, with a small discrepancy arising mainly from the growth of the $m
= \pm 1$ components in the numerical result due to the dynamical
instability.
The $m = 0$ component reaches a maximum at $t \simeq \ln(16 / |\nu|) / (2
|\alpha|)$ and the maximum value of $|\zeta_0(t)|^2$ is given by $\simeq
-A_0 / |\nu| \simeq 1 + 10 q / \tilde c_2$.

We note that a characteristic time scale of the dynamics in
Fig.~\ref{f:cyclic} is $\sim 100$ ms, which is much larger than the $\sim
10$ ms seen in Fig.~\ref{f:unifev}.
This is because the former time scale is set by the dynamical instability,
while the latter is set by the time evolution from a non-stationary
state.
This method can thus determine the value of $c_2$, provided that (i)
$c_2$ is positive, (ii) the magnetic field is suppressed so that $|q| <
\tilde c_2 / 10$, and (iii) the time scale of the dynamics is much faster
than the lifetime of the $f = 2$ state.
If $c_2$ is negative, the present method cannot determine the value of
$c_2$.

In the case of Fig.~\ref{f:cyclic} ($|q| < \tilde c_2 / 10$), the $m
= 0$ population initially increases for $c_2 > 0$, which is opposite to
the case of Fig.~\ref{f:unifev} ($|q| > |\tilde c_2| / 10$).
Nevertheless, the method to determine the sign of $c_2$ in
Sec.~\ref{ss:main} is still applicable, since we can easily find whether
we are in a region of $|q| > \tilde c_2 / 10$ or $|q| < \tilde c_2 / 10$.
If the amplitude of the initial oscillation monotonically decreases
without changing sign with an increase in the magnetic field, we can
conclude that $|q| > \tilde c_2 / 10$.

\section{Spin dynamics in the trapped system}
\label{s:trap}

We investigate the spin dynamics in a trap potential to verify the results
in Sec.~\ref{s:homo} in a realistic situation.
We use an axisymmetric trap potential with radial and axial frequencies
$(\omega_\perp, \omega_z) = 2 \pi \times (237, 21)$ Hz, which was used in
the experiment of Ref.~\cite{Kuwamoto}.
The two-body loss rate was found to be roughly independent of the magnetic
field and therefore it is considered to be insensitive to the spin-mixing
dynamics~\cite{Kuwamoto}.
We therefore take into account the two-body loss by adding the term $-i
\hbar K_2 |\psi_m|^2 \psi_m / 2$ to the right-hand side of
Eq.~(\ref{GP}) with $K_2 = 1.4 \times 10^{-13} {\rm cm}^3 {\rm s}^{-1}$.
We prepare the ground state $\psi_g$ for the $m = -2$ state by the
imaginary-time propagating method~\cite{Dalfovo} and transfer it to each
spin component as $\psi_0 = \sqrt{0.298} \psi_g$, $\psi_{\pm 1} =
\sqrt{0.001} \psi_g$, and $\psi_2 = \psi_{-2} = \sqrt{0.35} \psi_g$.
We use this state as an initial state, where the initial number of atoms
is assumed to be $N(t = 0) = 4.7 \times 10^5$~\cite{Kuwamoto}.

\begin{figure}[tb]
\includegraphics[width=8.4cm]{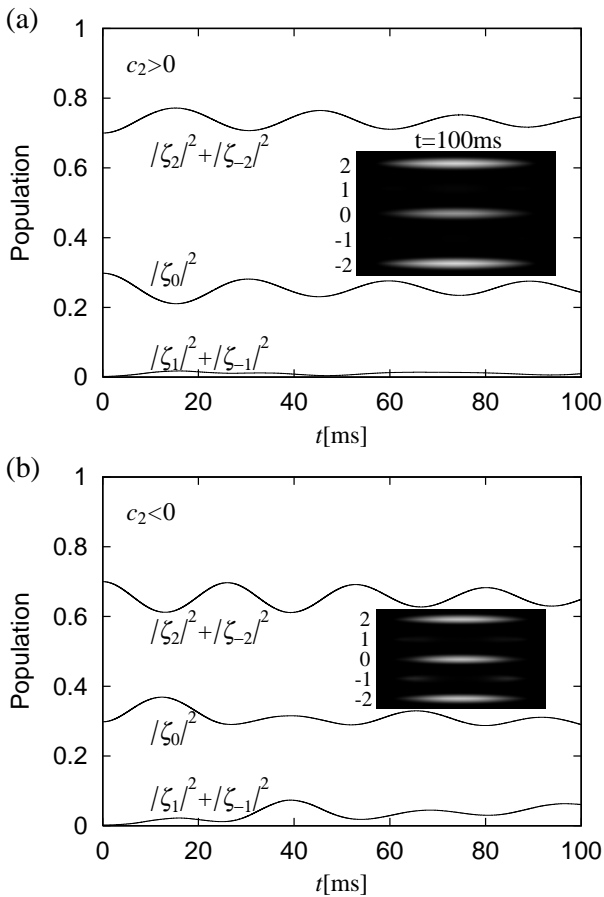}
\caption{
Time evolution of the relative population $\int d \bm{r} |\psi_m|^2 / N$
of each spin component $m$ in the trap with $\omega_\perp = 237$ Hz and
$\omega_z = 21$ Hz~\protect\cite{Kuwamoto}.
The scattering lengths are assumed to be $(c_0, c_1, c_2) M / (4 \pi
\hbar^2) = (100 a_{\rm B}, a_{\rm B}, a_{\rm B} )$ in (a) and $(100 a_{\rm
B}, a_{\rm B}, -a_{\rm B})$ in (b), where $a_{\rm B}$ is the Bohr radius.
The strength of the magnetic field is 250 mG, and the two-body loss
coefficient is $K_2 = 1.4 \times 10^{-13} {\rm cm}^3 {\rm
s}^{-1}$~\protect\cite{Kuwamoto}.
The $m = -2$ ground state $\psi_g$ is prepared, and is initially
transferred to each spin component as $\psi_0 = \sqrt{0.298} \psi_g$,
$\psi_{\pm1} = \sqrt{0.001} \psi_g$, and $\psi_2 = -\psi_{-2} =
\sqrt{0.35} \psi_g$.
The insets show the column density distributions of the $m = -2, \cdots,
2$ components from top to bottom at $t = 100$ ms, each having a $70 \times
8 \mu{\rm m}$ field of view.
}
\label{f:trap}
\end{figure}

The time evolution of each spin population obtained by numerically solving
the GP equation~(\ref{GP}) with the two-body loss terms is shown in
Fig.~\ref{f:trap}.
We find that the aforementioned dependence of the initial spin evolution
on the sign of $c_2$ is valid in the trapped system, i.e., the
quantity $|\zeta_0|^2 = \int d\bm{r} |\psi_0|^2 / N$ first decreases for
$c_2 > 0$ and increases for $c_2 < 0$.
The growth in $|\psi_{\pm 1}|^2$ due to the dynamical instability and the
two-body loss does not affect this dependence.
Thus, we have confirmed that the sign of $c_2$ can be determined from the
spin dynamics in a realistic situation.
We find that the growth of the $m = \pm 1$ components is larger for $c_2 <
0$ than for $c_2 > 0$, which is similar to the homogeneous case shown in
Fig.~\ref{f:unifev}.

The insets in Fig.~\ref{f:trap} show the column density distribution of
each spin component at $t = 100$ ms.
The initial distribution is almost preserved at $t = 100$ ms and no spin
domains are formed.

\section{Preparation of the initial spin state}
\label{s:exp}

As shown in Fig.~\ref{f:phase}, our method to determine the sign of $c_2$
hinges on the relative phase $\chi_2 + \chi_{-2} - 2 \chi_0$
in the initial state (\ref{initz}).
Therefore we must control the phase experimentally.
In the experiments of Refs.~\cite{Schmal,Kuwamoto}, the
rapid-adiabatic-passage technique~\cite{Mewes} was employed to prepare the
initial state.
This method can control the phase in the initial state in principle, if
the relative phases between the applied rf fields are controlled.
We propose here a new method to prepare the initial state using the Raman
transition.

\begin{figure}[tb]
\includegraphics[width=8.4cm]{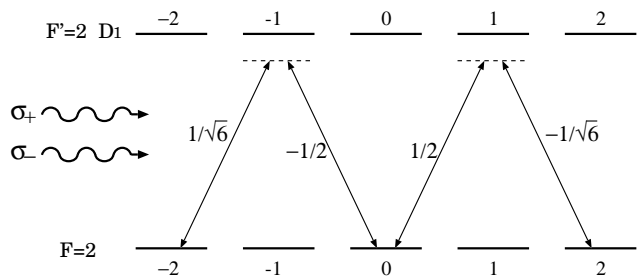}
\caption{
Schematic illustration of the energy levels of the $f = 2$ and $f' = 2$
$D_1$ states.
The circularly polarized light $\sigma_\pm$ changes $m$ by $\pm 1$ and
hence starting from the $f = 2$, $m = 0$ state, the system undergoes
transitions only to $f' = 2$, $m = \pm 1$ and $f = 2$, $m = \pm 2$ states
with the Clebsch-Gordan coefficients given by $\pm 1 / 2$ and $\mp 1 /
\sqrt{6}$, respectively.
}
\label{f:level}
\end{figure}

We consider the Raman transition, for example, between the $f = 2$ and $f'
= 2$ $D_1$ states by circularly polarized photons, as illustrated in
Fig.~\ref{f:level}.
The initial state is assumed to be the $m = 0$ lower state.
Hence, the relevant states are the $m = 0$, $\pm 2$ lower states and the
$m = \pm 1$ upper states, whose amplitudes are denoted by $\zeta_0$,
$\zeta_{\pm 2}$, and $\zeta_{\pm 1}'$, respectively.
The equations of motion are given by
\begin{subequations} \label{raman1}
\begin{eqnarray}
i \hbar \dot{\zeta}_{\pm 2} & = & (\pm 2 p + 4 q) \zeta_{\pm 2} \mp
\frac{1}{\sqrt{6}} g_\mp^* e^{i \omega_\mp t} \zeta_{\pm 1}', \\
i \hbar \dot{\zeta}_{\pm 1}' & = & (E_{D_1} \pm p' + q') \zeta_{\pm 1}'
\mp \frac{1}{\sqrt{6}} g_\mp e^{-i \omega_\mp t} \zeta_{\pm 2}
\nonumber \\ 
& & \pm \frac{1}{2} g_\pm e^{-i \omega_\pm t} \zeta_0,
\\
i \hbar \dot{\zeta}_0 & = & \frac{1}{2} g_+^* e^{i \omega_+ t} \zeta_1' -
\frac{1}{2} g_-^* e^{i \omega_- t} \zeta_{-1}',
\end{eqnarray}
\end{subequations}
where $g_\pm$ and $\omega_\pm$ are the coupling constants and frequencies
of the $\sigma_\pm$ photons, $E_{D_1}$ is the transition energy of the
$D_1$ line, and $p'$ and $q'$ are the linear and quadratic Zeeman energies
for the upper hyperfine manifold.
The coupling constants $g_\pm$ are given by the product of the amplitude
of the circularly polarized light and the dipole matrix element between
the $s$ and $p$ states.
Using the new variables $\tilde{\zeta}_{\pm 2} = e^{\pm 2ipt} \zeta_{\pm
2}$ and $\tilde{\zeta}_{\pm 1}' = e^{i \omega_\pm t} \zeta_{\pm 1}$, the
coefficients on the right-hand side of Eq.~(\ref{raman1}) become time
independent.
Assuming $\omega_\pm = E_{D_1} - \Delta \pm p + q'$ and $\Delta \gg |p -
p'|$, and eliminating $\zeta_{\pm 1}'$ from Eq.~(\ref{raman1}), we obtain
\begin{subequations} \label{raman2}
\begin{eqnarray}
i \hbar \dot{\tilde{\zeta}}_{\pm 2} & = & \left(4 q - \frac{|g_\mp|^2}{6
\Delta} \right) \tilde{\zeta}_{\pm 2} + i \hbar \Omega \zeta_0, \\
i \hbar \dot{\tilde{\zeta}}_0 & = & -\frac{1}{4 \Delta} \left( |g_+|^2 +
|g_-|^2 \right) \zeta_0 - i \hbar (\Omega^* \tilde \zeta_2 - \Omega \tilde
\zeta_{-2}),
\nonumber \\
\end{eqnarray}
\end{subequations}
where $i \hbar \Omega = g_+ g_-^* / (2 \sqrt{6} \Delta)$.
If the amplitudes of the $\sigma_\pm$ fields are the same, i.e., $|g_+| =
|g_-|$, and if the magnetic field is applied so that the condition $q =
-|g_+|^2 / (12 \Delta)$ is met, the coefficients of the first terms on the
right-hand side of Eq.~(\ref{raman2}) become equal, i.e., $4q - |g_\mp|^2
/ (6 \Delta) = -(|g_+|^2 + |g_-|^2) / (4 \Delta) = 6q$.
The solutions are thus obtained as
\begin{subequations}
\begin{eqnarray}
\zeta_2 & = & e^{-i (2p + 6q) t / \hbar} \frac{\Omega}{\sqrt{2
|\Omega|^2}} \sin \sqrt{2 |\Omega|^2} t, \\
\zeta_0 & = & e^{-i 6q t} \cos \sqrt{2 |\Omega|^2} t, \\
\zeta_{-2} & = & -e^{-i (-2p + 6q) t / \hbar} \frac{\Omega^*}{\sqrt{2
|\Omega|^2}} \sin \sqrt{2 |\Omega|^2} t.
\end{eqnarray}
\end{subequations}
We note that the phase difference is always $\chi = \chi_2 + \chi_{-2} - 2
\chi_0 = \pi$, which is independent of the phase of the circularly
polarized light and the applied magnetic field.
This method is thus suitable for preparing the initial state
(\ref{init}).

We can check whether the prepared state has the correct relative phase
$\chi = \pi$.
Rotating the spin state (\ref{initz}) by $\pi / 2$ around the axis on the
$x$-$y$ plane as $\bar{\bm{\zeta}} = e^{-i \pi S_\phi / 2} \bm{\zeta}$
with $S_\phi = S_x \cos \phi + S_y \sin \phi$, we find that the population
of each component becomes
\begin{subequations}
\begin{eqnarray}
\label{rotz}
|\bar{\zeta}_2|^2 & = & |\bar{\zeta}_{-2}|^2 = \frac{1}{8} \Bigl( 3 \cos^2
\theta - \sqrt{3} \sin 2\theta \cos \frac{\chi}{2} \cos \chi' \nonumber \\
& & + \sin^2 \theta \cos^2 \chi' \Bigr), \\
|\bar{\zeta}_1|^2 & = & |\bar{\zeta}_{-1}|^2 = \frac{1}{2} \sin^2 \theta
\sin^2 \chi', \\
|\bar{\zeta}_0|^2 & = & \frac{1}{4} \Bigl( \cos^2 \theta + \sqrt{3} \sin
2\theta \cos \frac{\chi}{2} \cos \chi' \nonumber \\
& & + 3 \sin^2 \theta \cos^2 \chi' \Bigr),
\end{eqnarray}
\end{subequations}
where $\chi' = 2 \phi + (\chi_2 - \chi_{-2}) / 2$.
Let us consider the quantity
\begin{equation} \label{fz2}
\bar{f}_z^2 = 2(4|\bar{\zeta}_2|^2 + |\bar{\zeta}_1|^2)
= 1 + 2 \cos^2 \theta - \sqrt{3} \sin 2\theta \cos \frac{\chi}{2} \cos
\chi',
\end{equation}
which can be measured by the spin-dependent non-destructive imaging
reported in Ref.~\cite{Higbie}.
We note that if $\chi = \pi$, Eq.~(\ref{fz2}) is independent of $\chi'$
and is constant, while if $\chi \neq \pi$, Eq.~(\ref{fz2}) oscillates,
since the phase $\chi_2 - \chi_{-2}$ rotates in the magnetic field.
We can therefore confirm that the desired initial state (\ref{init}) is
obtained if the result of the polarization-dependent phase-contrast
imaging~\cite{Higbie} is constant in time.

\section{Conclusions}
\label{s:conc}

We have proposed a method to experimentally determine the sign of the
singlet-pair energy of a spin-2 BEC from spin exchange dynamics, which
determines whether the $f = 2$ ${}^{87}{\rm Rb}$ BEC at zero magnetic
field is antiferromagnetic or cyclic.
We have obtained analytic solutions of the multicomponent GP
equations and found that if we prepare the initial state in the magnetic
sublevels $m = 0$, $\pm 2$ with the appropriate relative phase
relationship, we can determine the sign of $c_2$ from the sign of the
initial change in the spin populations.
For example, in the case of the initial state in Eq.~(\ref{initz}), we
can conclude that $c_2$ is positive or negative if the $m = 0$ component
first decreases or increases.
We can also determine the magnitude of $c_2$ from the amplitude of the
oscillation in the spin populations.

Since only the initial evolution for a period of $\sim 10$ ms is needed,
we can use this method in the presence of atom loss by inelastic
collisions and even in the presence of dynamical instabilities.
We have numerically confirmed that this method is applicable for a
trapped system with a realistic two-body loss coefficient $K_2$.
The required condition for the magnetic field is $|q| > \tilde c_2 / 10$,
so our method works even when the quadratic Zeeman energy exceeds the
singlet-pair energy.
The initial spin state in which the populations and relative phases 
are controlled as Eq.~(\ref{initz}) can be prepared by, e.g., the Raman
technique as shown in Sec.~\ref{s:exp}.

We have shown that the coherent spin dynamics serve as an efficient probe
for investigating spinor properties especially when the atomic species
have a short lifetime and therefore the equilibrium state cannot be
reached.
The coherent spin dynamics may also reveal other equilibrium spinor
properties as well as nonequilibrium spinor properties that cannot be
obtained from the static equilibrium.

\begin{acknowledgments}
We would like to thank T. Kuwamoto and T. Hirano for fruitful
discussions.
This work was supported by Grant-in-Aids for Scientific Research (Grant
No.\ 17740263 and No.\ 15340129) and by a 21st Century COE program at
Tokyo Tech ``Nanometer-Scale Quantum Physics,'' from the Ministry of
Education, Culture, Sports, Science and Technology of Japan.
\end{acknowledgments}

\appendix

\section{Derivation of the solution (\protect\ref{SOL1})}
\label{a:derivation}

In this appendix, we solve Eq.~(\ref{GP02}) for the initial condition
(\ref{initz}) with $\chi_0 = \chi_2 = 0$ and $0 \leq \theta \leq \pi / 2$,
and derive the solution (\ref{SOL1}).
Defining a new variable
\begin{equation} \label{z0}
z_0(t) \equiv e^{i (\tilde c_2 / 5 + 4 q) t / \hbar} \zeta_0(t),
\end{equation}
we can rewrite Eq.~(\ref{eqzeta0}) as
\begin{equation} \label{eqz0}
\ddot{z}_0 = P z_0 + Q |z_0|^2 z_0,
\end{equation}
where
\begin{eqnarray}
P & = & \frac{1}{\hbar^2} \left( \frac{2 \tilde c_2}{5} \varepsilon_s -
\frac{\tilde c_2^2}{25} - 16 q^2 - \frac{16 \tilde c_2 q}{5} \right), \\
Q & = & \frac{16 \tilde c_2 q}{5 \hbar^2},
\end{eqnarray}
with the spin-dependent energy
\begin{equation}
\varepsilon_s = \frac{\tilde c_2}{10} |2 \zeta_2 \zeta_{-2} + \zeta_0^2|^2
+ 4q (1 - |\zeta_0|^2).
\end{equation}

The initial conditions for $z_0$ are $z_0(0) = \zeta_0(0) = \cos\theta_0$
and
\begin{eqnarray} \label{zdot0}
\dot{z}_0(0) & = & \frac{i}{\hbar} \left( \frac{\tilde c_2}{5} + 4 q
\right) \zeta_0(0) \nonumber \\
&& - \frac{i \tilde c_2}{5 \hbar} \zeta_0(0) \left[ 2 e^{i
\chi_{-2}} \zeta_2(0)^2 + \zeta_0^2(0) \right],
\end{eqnarray}
where we have used Eq.~(\ref{GP02_0}).

We write the complex variable as $z_0(t) \equiv R(t) e^{i \phi(t)}$ with
real functions $R(t)$ and $\phi(t)$, where $\phi(0) = 0$ and $z_0(0) =
R(0)$.
Substituting this into Eq.~(\ref{eqz0}) and taking the real and imaginary
parts, we obtain
\begin{eqnarray}
\ddot{R} - P R - Q R^3 - R \dot{\phi}^2 & = & 0, \label{rphi1} \\
2 \dot{R} \dot{\phi} + R \ddot{\phi} & = & 0. \label{rphi2}
\end{eqnarray}
These equations can be integrated to give
\begin{eqnarray}
\dot{R}^2 & = & P R^2 + \frac{Q}{2} R^4 + C_R - \frac{C_\phi^2}{R^2},
\label{rdot2} \\
\dot{\phi} & = & \frac{C_\phi}{R^2}, \label{phidot}
\end{eqnarray}
where $C_R$ and $C_\phi$ are constants of integration.
The time derivative of $z_0$ at $t = 0$, $\dot{z}_0(0) = \dot{R}(0) + i
\dot{\phi}(0) R(0)$, gives ${\rm Re} \dot{z}_0(0) = \dot{R}(0)$ and ${\rm
Im} \dot{z}_0(0) = \dot{\phi}(0) R(0)$.
The constant $C_\phi$ is then written as
\begin{equation}
C_\phi = \dot{\phi}(0) R^2(0) = z_0(0) {\rm Im} \dot{z}_0(0).
\end{equation}
The constant $C_R$ is similarly obtained as
\begin{equation}
C_R = |\dot{z}_0(0)|^2 - P z_0^2(0) - \frac{Q}{2} z_0^4(0).
\end{equation}

We introduce a new variable $\rho$ through
\begin{equation} \label{rho}
R^2 = A_0 \rho^2 + B_0,
\end{equation}
where $A_0$ and $B_0$ are constants.
Substituting Eq.~(\ref{rho}) into Eq.~(\ref{rdot2}), we obtain
\begin{eqnarray} \label{rhoeq}
A_0^2 \dot{\rho}^2 & = & \frac{Q}{2} A_0^3 \rho^4 + \left( \frac{3Q}{2}
B_0 + P \right) A_0^2 \rho^2 \nonumber \\
& & + \left( \frac{3Q}{2} B_0^2 + 2 P B_0 + C_R \right) A_0 \nonumber \\
& & + \left( \frac{Q}{2} B_0^3 + P B_0^2 + C_R B_0 - C_\phi^2 \right)
\rho^{-2}. \nonumber \\
\end{eqnarray}
We determine $B_0$ so that the $\rho^{-2}$ term in Eq.~(\ref{rhoeq})
vanishes:
\begin{equation} \label{eqb0}
\frac{Q}{2} B_0^3 + P B_0^2 + C_R B_0 - C_\phi^2 = 0.
\end{equation}
The right-hand side (rhs) of Eq.~(\ref{rhoeq}) then becomes quartic with
respect to $\rho$.
We determine $A_0$ so that the rhs of Eq.~(\ref{rhoeq}) becomes
proportional to $(1 - \rho^2)(1 - \nu \rho^2)$; the result is
\begin{eqnarray} \label{a0org}
A_0 & = & \frac{1}{2Q} \Bigl[ -3 Q B_0 - 2 P \nonumber \\
& & \pm \sqrt{-(3 Q B_0 - 2 P) (Q B_0 + 2 P) - 8 Q C_R} \Bigr] \nonumber
\\
\end{eqnarray}
with
\begin{equation} \label{nuorg}
\nu = -\frac{Q A_0}{Q A_0 + 2 P + 3 Q B_0}.
\end{equation}
Equation (\ref{rhoeq}) thus becomes
\begin{equation}
\dot{\rho}^2 = \frac{Q A_0}{2 \nu} (1 - \rho^2) (1 - \nu \rho^2).
\end{equation}
This equation can be integrated to give
\begin{equation}
\int_{\rho(0)}^{\rho(t)} \frac{d\rho}{\sqrt{(1 - \rho^2) (1 - \nu
\rho^2)}} = \sqrt{\frac{Q A_0}{2 \nu}} t.
\end{equation}
We rewrite this equation as
\begin{equation} \label{intrho}
\int_{0}^{\rho(t)} \frac{d\rho}{\sqrt{(1 - \rho^2) (1 - \nu
\rho^2)}} = \alpha t + \beta_0,
\end{equation}
where
\begin{eqnarray}
\alpha & = & \sqrt{\frac{Q A_0}{2 \nu}}, \label{alphorg} \\
\beta_0 & = & \int_{0}^{\rho(0)} \frac{d\rho}{\sqrt{(1 - \rho^2) (1 - \nu
\rho^2)}}. \label{beta0org}
\end{eqnarray}
The left-hand side of Eq.~(\ref{intrho}) is an elliptic integral of the
first kind and hence $\rho(t) = {\rm sn}(\alpha t + \beta_0 |
\nu)$~\cite{Handbook}.
Thus,
\begin{equation} \label{zeta0}
|\zeta_0(t)| = R(t) = \sqrt{A_0 {\rm sn}^2(\alpha t + \beta_0 | \nu) +
 B_0}.
\end{equation}
From Eqs.~(\ref{z0}) and (\ref{phidot}), the phase $\varphi_0$ of
$\zeta_0$ is given by
\begin{equation} \label{phi0}
\varphi_0(t) = -\frac{1}{\hbar} \left( \frac{\tilde c_2}{5} + 4 q \right)
t + \int_0^t d\tau \frac{C_\phi}{A_0 {\rm sn}^2(\alpha t + \beta_0 | \nu)
+ B_0}.
\end{equation}

Similarly, we can obtain the solution for $\zeta_{\pm 2}(t)$.
From the form of Eq.~(\ref{GP02}) and the initial condition $\zeta_{-2}(0)
= e^{i \chi_{-2}} \zeta_2(0)$, we find that the relation $\zeta_{-2}(t)
= e^{i \chi_{-2}} \zeta_2(t)$ always holds.
Hence, we consider only $\zeta_2(t)$.
Eliminating $\zeta_0(t)$ from Eq.~(\ref{GP02}), we have the equation of
motion for $\zeta_2(t)$ as
\begin{equation} \label{ddotz2}
\hbar^2 \ddot{\zeta}_2 = \left[ \frac{2 \tilde c_2}{5} (\varepsilon_s + 4
q) - 16 q^2 - \frac{32 \tilde c_2 q}{5} |\zeta_2|^2 \right] \zeta_2 - i
\frac{2 \tilde c_2}{5} \hbar \dot{\zeta}_2.
\end{equation}
The new variable $z_2(t) = e^{i \tilde c_2 t / (5 \hbar)} \zeta_2(t)$
reduces Eq.~(\ref{ddotz2}) to
\begin{equation} \label{z2eq}
\ddot{z}_2 = P' z_2 + Q' |z_2|^2 z_2,
\end{equation}
where $P' = (2 \tilde c_2 \varepsilon_s / 5 - \tilde c_2^2 / 25 - 16 q^2 +
8 \tilde c_2 q / 5) / \hbar^2$ and $Q' = -32 \tilde c_2 q / (5 \hbar^2)$.
The initial conditions for $z_2$ are $z_2(0) = \zeta_2(0) = \sin\theta_0 /
\sqrt{2}$ and
\begin{equation} \label{zdot2}
\dot{z}_2(0) = \frac{i \tilde c_2}{5 \hbar} \zeta_2(0) \left[ 1 - 2
\zeta_2^2(0) - e^{-i \chi_{-2}} \zeta_0^2(0) \right] - \frac{4 i q}{\hbar}
\zeta_2(0).
\end{equation}
Equation (\ref{z2eq}) has the same form as Eq.~(\ref{eqz0}), in which $P$
and $Q$ are replaced by $P'$ and $Q'$, and therefore the derivation from
Eq.~(\ref{rphi1}) to Eq.~(\ref{phi0}) can similarly be applied.

We consider the case of $\chi_{-2} = \pi$ in the initial condition given
in Eq.~(\ref{initz}).
In this case, Eq.~(\ref{zdot0}) is pure imaginary, and Eq.~(\ref{eqb0})
can be solved to give $B_0 = z_0^2(0) = \zeta_0^2(0)$.
Equation (\ref{a0org}) then reduces to
\begin{equation} \label{appa0}
A_0 = \left\{ \begin{array}{l}
\displaystyle
1 - 2 \zeta_0^2(0) + \frac{10 q}{\tilde c_2}, \\
\displaystyle
\frac{\tilde c_2}{10 q} \zeta_0^2(0) \left[ 1 - \zeta_0^2(0) \right],
\end{array} \right.
\end{equation}
where the upper and lower expressions correspond to the plus and minus
signs of $\pm$ in Eq.~(\ref{a0org}).
Using this $A_0$, $\nu$ and $\alpha$ become
\begin{eqnarray}
\label{appnu} 
\nu & = & \left\{ \begin{array}{l}
\displaystyle \frac{10 q \left\{ \tilde c_2 \left[ 1 - 2 \zeta_0^2(0)
\right] + 10 q \right\}}{\displaystyle \tilde c_2^2 \zeta_0^2(0) \left[ 1
- \zeta_0^2(0) \right]}, \\
\displaystyle \frac{\tilde c_2^2 \zeta_0^2(0) \left[ 1 - \zeta_0^2(0)
\right]}{\displaystyle 10 q \left\{ \tilde c_2 \left[ 1 - 2 \zeta_0^2(0)
\right] + 10 q \right\}},
\end{array} \right. \\
\label{appalph}
\alpha & = & \left\{ \begin{array}{l}
\displaystyle
\frac{2 |\tilde c_2|}{5 \hbar} \sqrt{\zeta_0^2(0) \left[ 1 - \zeta_0^2(0)
\right]}, \\
\displaystyle
\sqrt{\frac{8q}{5 \hbar^2} \left\{ \tilde c_2 \left[ 1 - 2 \zeta_0^2(0)
\right] + 10 q \right\}}.
\end{array} \right.
\end{eqnarray}
It follows from Eq.~(\ref{rho}) and $B_0 = z_0^2(0)$ that $\rho(0) = 0$
and hence $\beta_0 = 0$.
Using the relation
\begin{equation}
{\rm sn}(\alpha t | \nu) = \frac{1}{\sqrt{\nu}} {\rm sn}(\sqrt{\nu} \alpha
t | \nu^{-1}),
\end{equation}
we can show that the two expressions in Eqs.~(\ref{appa0}), (\ref{appnu}),
and (\ref{appalph}) are equivalent to each other.
We used the lower expressions in Sec.~\ref{s:homo}.
The $m = \pm 2$ components $\zeta_2(t) = -\zeta_{-2}(t)$ have the same
form as Eq.~(\ref{zeta0}), in which $A_0$, $\beta_0$, and $B_0$ are
replaced by $A_2$, $\beta_2$, and $B_0$.
These constants are given by
\begin{equation}
A_2 = \left\{ \begin{array}{l} \displaystyle
\frac{1}{2} - 2 \zeta_2^2(0) - \frac{5 q}{\tilde c_2}, \\
\displaystyle
-\frac{\tilde c_2}{10 q} \zeta_2^2(0) \left[ 1 - 2 \zeta_2^2(0) \right],
\end{array} \right.
\end{equation}
$\beta_2 = 0$, and $B_2 = \zeta_2^2(0)$.

\begin{widetext}
Finally, we consider the case of $\chi_{-2} = 0$ in the initial
condition (\ref{initz}), i.e., all the components are real and positive.
We can solve Eq.~(\ref{eqb0}) also in this case to give $B_0 = z_0^2(0)$
and hence $\beta_0 = 0$.
The constants $A_0$, $\nu$, and $\alpha$ reduce to
\begin{eqnarray}
\label{samea0}
A_0 & = & \frac{1}{2 \tilde c_2 q} \left\{ q \left\{ \tilde c_2 [1 - 2
\zeta_0^2(0)] + 10 q \right\} - |q| \sqrt{(\tilde c_2 + 10 q)^2 - 40
\tilde c_2 q \zeta_0^2(0)} \right\},
\nonumber \\
\\
\label{samenu}
\nu & = & \frac{q \left\{ \tilde c_2 [1 - 2 \zeta_0^2(0)] + 10 q \right\}
- |q| \sqrt{(\tilde c_2 + 10 q)^2 - 40 \tilde c_2 q \zeta_0^2(0)}}{q
\left\{ \tilde c_2 [1 - 2 \zeta_0^2(0)] + 10 q \right\} + |q|
\sqrt{(\tilde c_2 + 10 q)^2 - 40 \tilde c_2 q \zeta_0^2(0)}}, \\
\label{samealph}
\alpha & = & \sqrt{\frac{8 \tilde c_2 q A_0}{5 \hbar^2 \nu}},
\end{eqnarray}
where we take the lower sign in Eq.~(\ref{a0org}).
Since $\left\{ \tilde c_2 [1 - 2 \zeta_0^2(0)] + 10 q \right\}^2 < (\tilde
c_2 + 10 q)^2 - 40 \tilde c_2 q \zeta_0^2(0)$, we find $\tilde c_2 A > 0$
and $\nu < 0$.
\end{widetext}

\end{document}